\documentclass[aip,reprint,onecolumn]{revtex4-1}
\usepackage{graphicx}

\begin{document}

\title{Femtosecond single- to few-electron point-projection microscopy for nanoscale dynamic imaging}

\author{A. R. Bainbridge}
\affiliation{Department of Physics, College of Science, Swansea University, Singleton Park, Swansea, SA2 8PP, UK}
\affiliation{Accelerator Science and Technology Centre, STFC Daresbury Laboratory, Daresbury
 Science and Innovation Centre, Keckwick Lane, Daresbury, Cheshire, WA4 4AD, UK}

\author{C. W. Barlow-Myers}
\affiliation{Department of Physics, College of Science, Swansea University, Singleton Park, Swansea, SA2 8PP, UK}

\author{W. A. Bryan}
\email{w.a.bryan@swansea.ac.uk}
\affiliation{Department of Physics, College of Science, Swansea University, Singleton Park, Swansea, SA2 8PP, UK}

\date{\today}

\begin{abstract}
Femtosecond electron microscopy produces real-space images of matter in a series of ultrafast snapshots. Pulses of electrons self-disperse under space-charge broadening, so without compression, the ideal operation mode is a single electron per pulse. Here, we demonstrate for the first time femtosecond single-electron point projection microscopy (fs-ePPM) in a laser-pump fs-e-probe configuration. The electrons have an energy of only 150 eV and take tens of picoseconds to propagate to the object under study. Nonetheless, we achieve a temporal resolution with a standard deviation of 120 fs, combined with a spatial resolution of 100 nm, applied to a localized region of charge at the apex of a nanoscale metal tip induced by 30 fs 800 nm laser pulses at 50 kHz. These observations demonstrate real-space imaging of reversible processes such as tracking charge distributions is feasible whilst maintaining femtosecond resolution. Our findings could find application as a characterization method, which, depending on geometry could resolve tens of femtoseconds and tens of nanometres. Dynamically imaging electric and magnetic fields and charge distributions on sub-micron length scales opens new avenues of ultrafast dynamics. Furthermore, through the use of active compression, such pulses are an ideal seed for few-femtosecond to attosecond imaging applications which will access sub-optical cycle processes in nanoplasmonics. 

\end{abstract}


\maketitle

\section{Introduction}
A variety of fundamental physical, chemical and biological processes occur on ultrafast timescales and over length scales ranging from microns to {\AA}ngstroms. Direct visualization of such processes requires imaging with temporal and spatial resolution. Pump-probe experiments serve well in the time domain, whereby both the pump and probe pulses have durations shorter than the characteristic timescale of the process under study. There is a disparity between the ultimate goal of atomic scale imaging and the wavelength of visible and IR ultrafast laser sources, hence to approach the few nanometre to tens of picometre length-scale, current ultrafast imaging paradigms have turned to the diffraction of very short wavelength (x-ray) photons. X-ray Free Electron Lasers (X-FELs)\cite{xfel} enable diffraction in the femtosecond regime, facilitating the real-time observation of geometry modification of a protein \cite{Tenb}, laser-induced lattice dynamics in nanoparticles\cite{Clark}, or tracking charge density waves\cite{Moore}. 

Electrons have a de Broglie wavelength three orders of magnitude smaller than a photon of equivalent energy, the realization of which led to the development of transmission electron microscopy\cite{knollruska}. Producing electrons with energies from hundreds of eV to tens of keV is, as compared to XFELs, relatively straightforward. Combining femtosecond laser technology with electron sources, as first demonstrated with picosecond resolution by Zewail and co-workers\cite{ihee1}, followed by femtosecond structural dynamics resolved by Miller and co-workers\cite{miller}, introduced a novel approach to ultrafast imaging. The ongoing evolution of this field is producing unprecedented insights into fundamental processes influencing many research themes\cite{siwickgraphene, dudek, dwyerreview, gulde, ischenko, baum1, thinfilms, srireview}. 

Two emission processes have been demonstrated to produce ultrafast pulses of electrons making use of either the energy of one or more photons, or the resulting electric field induced by a flux of photons. Illuminating a planar photocathode with a laser pulse which exceeds the work function liberates electrons. Recent examples by Wann and co-workers\cite{Rob1, Rob2} indicate typical performance in the UV, Baumert and co-workers demonstrated a compact diffractometer with novel magnetic lens\cite{gerbig} and a two-photon apparatus by Ernstorfer and co-workers\cite{ernstorfer2} shows high temporal resolution up to 100 keV. Very high coherence electron pulses have been generated by Luiten and co-workers\cite{luiten1, luiten2, luiten3} and Scholten and co-workers\cite{McCull1, McCull2} by illuminating a cloud of ultracold atoms with a photon energy just above the ionization potential resulting in low temperature electrons, albeit at pulse durations limited by the narrow bandwidth employed.

The emission process used in the present work is inspired by the work of Ropers\cite{gulde, ropers, herink}, Ernstorfer\cite{ernstorfer1}, Hommelhoff\cite{schenk, hommelhoff, homelhoff2, kruger}, Baum \cite{Hoff, Baum2} and Barwick\cite{BarwickPPM}. A Nanoscale Metal Tip (NSMT) is a cylindrically symmetric wire that tapers to an apex with a radius of curvature (RoC) of tens of nanometres. Rather than using the photoelectric effect, the small RoC causes a significant field enhancement when an electric field is applied to the NSMT. This process has been studied for some time as DC field emission, whereby the electric field at the apex is sufficient to allow electrons near the Fermi level to tunnel directly into the continuum\cite{longchamp1, longchamp2}. Applying a femtosecond laser pulse of moderate intensity causes the same process\cite{batelaan} in the strong-field regime. Tunnelling is only possible during the laser pulse, hence the emitted electron pulse can be tens of femtoseconds in duration. 

In this work, we demonstrate a novel application of the simplest real-space ultrafast electron imaging method. DC point projection microscopy as pioneered by Fink and co-workers\cite{Tati}, allows straightforward magnification and holography. Femtosecond electron point-projection microscopy (fs-ePPM) has been demonstrated previously by Ropers\cite{gulde}, Ernstorfer\cite{ernstorfer1, Mull16} and Barwick\cite{BarwickPPM}, has advantages over more complex, higher energy configurations, namely being lensless and very compact. A laser-illuminated NSMT is a near-point source of electrons which diverge electrostatically, providing magnification. In fs-ePPM, magnifications of 5000 have been achieved \cite{BarwickPPM}, compared to the work of Fink and co-workers with DC field emission PPM\cite{longchamp1, longchamp2}, who have achieved magnifications of in excess of 10$^4$. Clearly, spatial resolution must be considered alongside temporal resolution: in the purely fs-ePPM experiments, Barwick\cite{BarwickPPM} presented a spatial resolution of 100 nm with an autocorrelation at source of 100 fs, and Ernstorfer\cite{ernstorfer1} demonstrated a temporal resolution of approximately 250 fs and a spatial resolution better than 200 nm. Ropers and co-workers employed a defocused mode of their femtosecond electron diffraction instrument showing a temporal resolution of 2 ps (FWHM), and the recent observations of Ernstorfer\cite{Mull16} show a spatial resolution below 100 nm, and a predicted temporal resolution of between 10 and 20 fs, albeit without demonstration at this stage. Nonetheless, this very recent work using a grating-coupled NSMT will allow the fs-ePPM magnification to exceed 10$^4$, with spatial resolutions well below 100 nm and temporal resolutions around 10 fs expected. 

Femtosecond electron emission from a planar photocathode, ultracold atoms or a NSMT generally results in a kinetic energy below 5 eV and a bandwidth defined by the laser pulse spectrum\cite{us_vmi, batelaan, herink, kruger}. Acceleration is necessary to minimize the kinematic dispersion due to variations in particle velocity. Furthermore, if there is more than one electron per pulse, the Coulomb repulsion results in temporal stretching, referred to as space-charge broadening\cite{Siwick, spacecharge1, spacecharge2}. Both kinematic dispersion and space-charge broadening are minimized by transporting the electrons to the target as quickly as possible\cite{Hoff}. Generating the shortest pulses in fs-ePPM requires as few electrons per pulse as possible, hence to generate real-space images with sufficient signal to perform time resolved measurements requires a process that can be cycled repeatedly, preferably at high laser repetition rates. Furthermore, we require a target which can be pumped with an infrared laser pulse and which reacts by causing a highly localized emission of charge, creating a measurable distortion to our femtosecond electron (fs-e) pulse. We therefore select a second NSMT as the target. 

The realization of this time-varying field-enhancement in tungsten, gold and silver has initiated the highly active research field of ultrafast nanophotonics or femtosecond nanophysics. Themes including optical near-field sensing, sub-cycle carrier-envelope phase phenomena, investigations into surface plasmons and laser-driven electron accelerators have emerged \cite{attonano1, attonano2, attonano3, attonano4}. The present work aligns itself to these efforts by looking to better quantify the spatial and temporal characteristics of femtosecond laser-driven electron microscopy to demonstrate a potential characterization method, and point to the potential of dynamic charge imaging. 

\begin{figure}
	\includegraphics[width=0.5\linewidth]{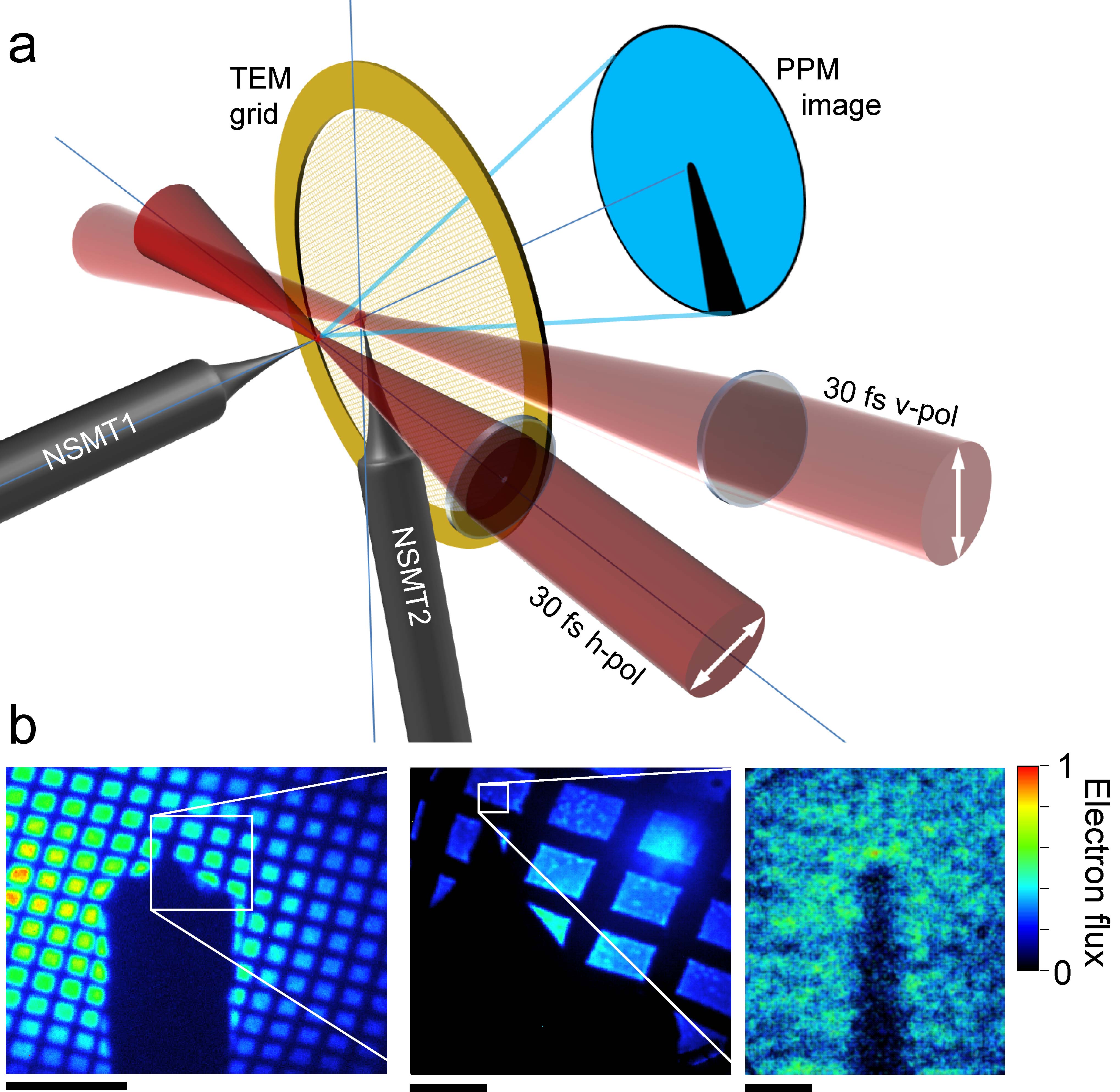}
	\caption{(a) Schematic of the femtosecond electron point-projection microscope. A nanoscale metal tip (NSMT1) was illuminated with horizontally polarized $\lambda$ = 800 nm, 30 fs laser pulses at a repetition rate of 50 kHz. The apex of NSMT1 had a radius of curvature of approximately 50 nm and the laser pulses were transmission focused to a spot size of 5.4 $\mu$m. A pulse energy of 45 nJ produced a peak intensity of 1.2 $\times$ 10$^{12}$ Wcm$^{-2}$ however the apex of NSMT1 experienced less than 1/20th of this intensity. NSMT1 was held at a potential of -150 V with respect to a grounded TEM grid, accelerating the resultant femtosecond electron (fs-e) pulses towards a distant microchannel plate detector, forming a point-projection image. Nanoscale metal tip 2 (NSMT2) was illuminated by vertically polarized laser pulses from the same source and a comparable intensity. A significantly longer focusing optic (\textit{f} = 300 mm as compared to \textit{f} = 75 mm) illuminates a region above the apex of NSMT2, resulting in a localized time-varying charge distribution. (b) Low magnification to high magnification PPM images of NSMT2 using electrons emitted from NSMT1. Each image is the sum of 5 $\times$ 10$^5$ laser shots, and the colour scale is the normalized electron flux. From left to right, the scale bars are 250 $\mu$m, 50 $\mu$m and 250 nm, and corresponding maximum fluxes of 8 $\times$ 10$^5$, 4 $\times$ 10$^5$ and 2 $\times$ 10$^4$ electrons.}
	\label{fig1}
\end{figure}

\section{Experimental}
An electron point-projection microscope was constructed from two electrochemically etched tungsten NSMTs, fabricated from polycrystalline tungsten wire of diameter 0.25 mm. This material is selected for its relatively low plasmonic response to the laser field\cite{Girard, Thomas15} facilitating fine control of the electron flux, and is mechanically hard with a high melting point, making it resistant to laser damage. Apexes of tens to hundreds of nanometres are produced using standard techniques, characterised using the Fowler-Nordheim method. For illustration, SEM images of typical NSMTs are to be found the literature. An illustration of the experimental configuration is presented in figure \ref{fig1}(a). A source tungsten NSMT (referred to as NSMT1 throughout) is the photocathode, positioned on a vibration-isolating mount located concentrically with a grounded TEM grid. The NSMT1-TEM grid distance was adjustable via an external translation stage. The TEM grid was in turn supported by a vibration isolating and electrically insulating mount, concentric to an electron detector 0.44 m further along the flight path. The detector consists of a pair of micro-channel plates (Burle) and a P22 phosphor screen (Kimball Physics), referred to as MCP+PS, and imaged by a CCD camera (Pike-145B, Allied Vision Technologies). A second tungsten nanotip (NSMT2) was placed between NSMT1 and the TEM grid, attached to the grid mount but with an independently controllable voltage. Independent control of the potential applied to NSMT2 and the TEM grid gave control over the electrostatic lensing of the probe electrons. 

The apparatus illustrated in figure \ref{fig1}(a) was contained in an ultra-high vacuum chamber with a base pressure of 3 $\times$ 10$^{-9}$ mbar. The chamber was configured such that both the tip and sample are secured directly to the optical table. A mu-metal shield surrounding the flight path between the TEM grid and the detector prevented disruption by magnetic fields, and a 3-axis Helmholtz coil neutralized magnetic fields in the interaction region and provided beam steering at low magnifications.

Femtosecond laser pulses from a Light Conversion Pharos system at a repetition rate of 50 kHz pumped a Non-collinear Optical Parametric Amplifier (NOPA) (Orpheus-N, Light Conversion) operating at a centre wavelength of 800 nm. The NOPA generated sufficient bandwidth to give a transform-limited duration of 14 fs, and was compressed to sub-30 fs in a prism compressor compensating a 3.3 mm thick fused silica window and 2.4 mm thick UV fused silica lens. The duration of the pulse was confirmed as 28 fs using Frequency-Resolved Optical Gating (FROG), providing shot-to shot monitoring.

The NOPA output was split 20:80 and each arm directed through separate periscopes configured to ensure that the light arriving at NSMT1 and NSMT2 was polarized along the tip axes. As the strong-field ejection of electrons from NSMTs has been demonstrated to be polarization dependent\cite{Yana,yana2}, illuminating with crossed polarizations ensured NSMT1 would not emit under illumination by the NSMT2 laser focus and vice versa. The pulse to NSMT2 passed through a delay stage introducing 500 ps of additional flight time with 6.7 fs precision.

Low, medium and high magnification fs-ePPM images are shown in figure \ref{fig1}(b), indicating how NSMT2 is located. Manipulating the magnification requires the distance between NSMT1 and NSMT2 to be varied, hence the position of the focus of the horizontally polarized laser pulse must move with NSMT1. 

\begin{figure}
	\includegraphics[width=0.5\linewidth]{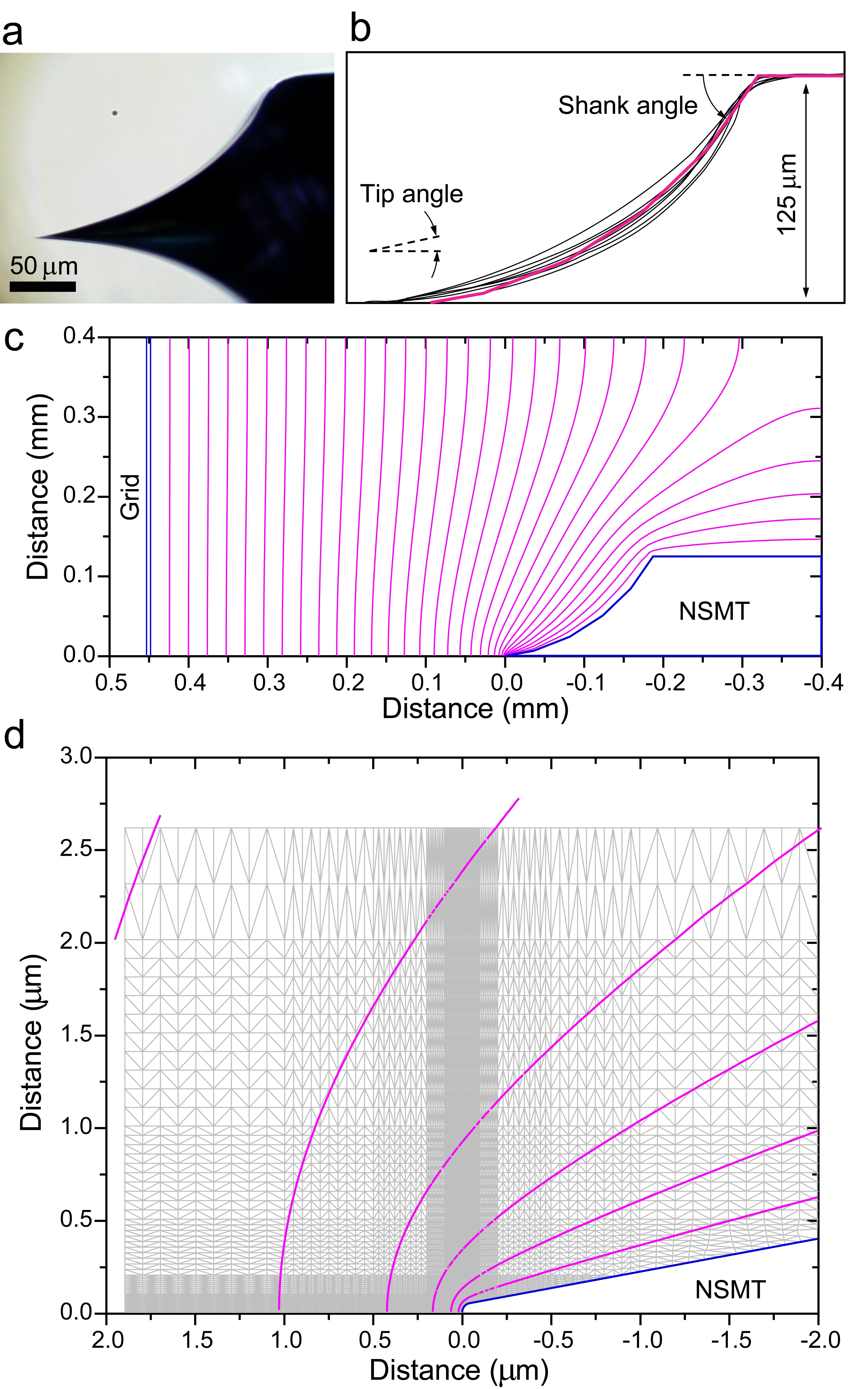}
	\caption{(a) Optical microscope image of a typical NSMT. (b) By thresholding and edge detection, a series of lines representing the typical shape of a NSMT is shown (black lines). The magenta line is the approximation to this shape used in the Superfish Poisson solver. (c) Superfish output of PPM geometry with the TEM grid at 0 V and the NSMT at -150 V. (d) Superfish field map at NSMT1, showing changing mesh size to accommodate millimetre flight length and nanometric tip apex. In (c) and (d), equipotentials are spaced by 10 V.}
	\label{fig2}
\end{figure}

\section{Modelling}
The Superfish Poisson Solver (LANL) was used to model the electrostatic field between NSMT1 and the plane of the TEM grid. The shape of a series of NSMTs was recorded using an optical microscope, shown in figure \ref{fig2}(a). Clearly, optical methods cannot resolve the apex of a NSMT, however they allow a typical shape to be determined, shown in \ref{fig2}(b). Average tip and shank angles were found to be 17 and 55 degrees. We approximated this combined profile with ten straight sections from the cylindrical edge of the tungsten wire, followed by a 20 $\mu$m section at the tip angle. The apex of the NSMT was a 50 nm RoC meeting this straight section at a tangent. With NSMT1 450 $\mu$m from the TEM grid, Superfish produced the representative electric field map shown in figure \ref{fig2}(c), with NSMT1 at -150V, and the TEM grid at 0 V. The influence of the small RoC of the apex is apparent from the bunching of field equipotentials. As the Poisson solver operates on a rectangular mesh, we adjusted the mesh spacing in the vicinity of the apex, as shown in figure \ref{fig2}(d).  

The Superfish electrostatic field was imported into the General Particle Tracer (GPT, Pulsar Physics)\cite{GPT}. NSMT2 is omitted from the calculation as it is known from the experimental results that it is possible to hold it at a voltage which prevents electrostatic lensing. This is also necessary to approximate a rotationally symmetric system, dramatically reducing computational requirements. As the number of electrons per pulse was initially unknown, our simulations of propagation from NSMT1 to the TEM grid requires the inclusion of space charge. GPT decouples the charge per pulse from number of simulation elements by defining identical macro-particles, here 50,000. The total charge of these macro-particles can be varied arbitrarily from zero. GPT solves Poisson's equation in three dimensions in the rest frame of the electron pulse for all macro-particles using a particle-in-cell method. 

The dynamic propagation of the fs-e pulse is modelled in GPT by assuming an emission energy of 4.5 eV and a bandwidth of 0.3 eV, in agreement with recent studies\cite{us_vmi, kruger, schenk}. The origin of the electrons was defined as the surface of the apex of NSMT1, and the emission time defined as a Gaussian distribution with a standard deviation (SD) of 30 fs. GPT calculates the trajectory of the electrons with a 5th-order Runge-Kutta algorithm with variable step size defined by an accuracy of 1:10$^{-6}$. 

\section{PPM magnification and source size}
\begin{figure}
	\includegraphics[width=0.4\linewidth]{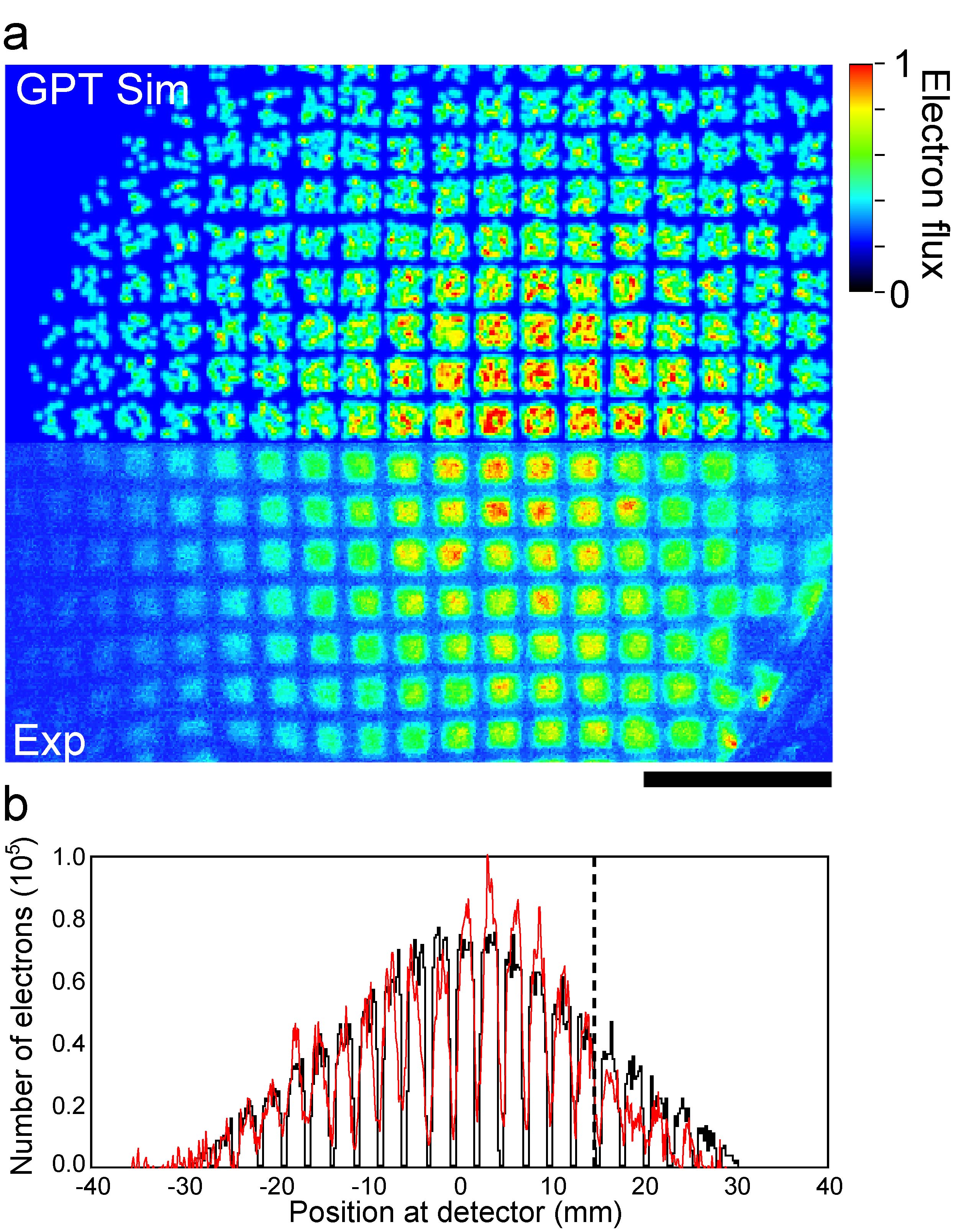}
	\caption{(a) Comparison of output of the GPT model (top) with experimental observations (bottom) of electron propagation from NSMT1, through a TEM grid at 2.45 mm from NSMT1, then on through a field free region to the detector 0.44 m distant. This agreement allows independent confirmation of magnification and source size. The scale bar is 350 $\mu$m at the TEM grid, and the colour scale is the normalized electron flux, where unity is equivalent to 10$^5$ electrons. (b) A direct comparison of the predicted and observed electron distribution following low magnification fs-ePPM. The agreement to the right of the dashed line is compromised by edge effects in the detector.}
	\label{fig3}
\end{figure}

With NSMT2 lowered out of the electron path, a fs-ePPM image of the TEM grid was recorded as in figure \ref{fig3}(a). The magnification was intentionally set low  so as to illuminate many grid squares. The TEM grid was 300-mesh copper, with a bar width of 25 $\mu$m and a centre to centre spacing of 85 $\mu$m. The GPT code was modified to include an analogous pattern of electron absorbing elements, representing the TEM grid. After propagating from NSMT1 to the TEM grid, the electron pulse then flies field-free to the detector 0.44 m distant and the distribution of arrival positions recorded. 

The spatial distribution of the initial electron position on the apex of NSMT1 was Gaussian, centred on-axis with a SD defined as a ratio of the NSMT1 apex radius. The NSMT1-TEM grid distance was varied in the simulation until the magnification of the TEM grid period matched that observed on the detector. The NSMT1 to TEM grid distance retrieved using this method was 2.45 $\pm$ 0.02 mm, with the estimate of uncertainty the minimum resolvable discrepancy between model and measurement. All subsequent translations of NSMT1 were made with respect to this fiducial. 

The width of the spatial distribution was then varied to best reproduce the experimental observations. Decreasing the SD of the Gaussian emission distribution from the NSMT1 apex (i.e 50 nm) in 5 nm steps, it was found that the best reproduction of the observations was found with an emission site of radius 10 nm. This is shown in figure \ref{fig3}(a) and a section through the middle of the observed distribution overlapped directly with the model result is shown in figure \ref{fig3}(b). A variation of 3 nm is the lower limit of observable deterioration in this agreement.  

\section{Electron flux calibration}
Quantifying and minimizing space charge broadening is one of the key aims of this work, hence an independent calibration of the fs-e flux is required. If the quantum efficiency of the apparatus were well known, direct calibration would in principle be possible, however MCP and PS performance varies as a function of time. 

The typical operating pressure of our MCP+PS was 6 $\times$ 10$^{-9}$ mbar, hence the detector is intrinsically low noise. Typically, the full active areas of the MCP+PS registered less than one background count per second with the fs-e source off. Under typical operating conditions, the front of the MCP pair was negatively biased by between 10 and 20 V to prevent stray electrons from our full-range pressure gauge (HPT 100, Pfeiffer Vacuum) being collected while still allowing fs-ePPM electrons to pass. 

We performed a simple calibration: with NSMT1 not emitting and the MCP+PS set to operational conditions, we varied the voltage on the front of the MCP until the detector was collecting tens of gauge electrons per exposure (286 ms). By manually counting electrons per exposure and summing for hundreds of exposures, we have a route to converting signal at the AVT Pike to electrons arriving at the front of the MCP. The variation of MCP efficiency with electron energy is known (80 at 300 eV to 50 at tens of eV), hence a count rate conversion was applied. This process was found to be reproducible day-to-day on the condition that the laser illumination of NSMT1 was held as constant, made possible by imaging the NSMT1 laser drive pulse on exit of the vacuum chamber.

\begin{figure}
	\includegraphics[width=0.5\linewidth]{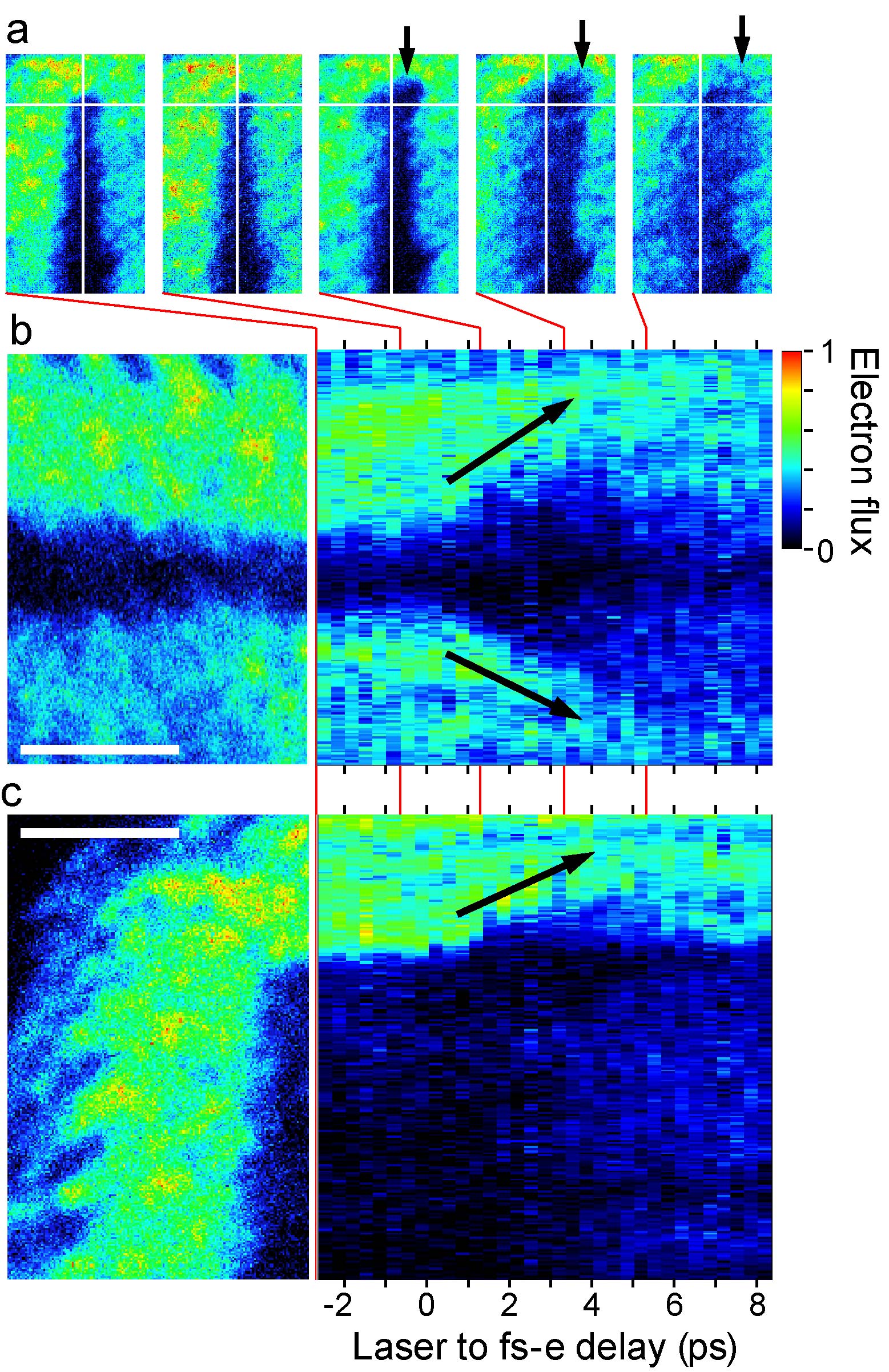}
	\caption{Femtosecond electron point-projection microscopy of nanoscale metal tip 2 (NSMT2) illuminated with an infrared 30 fs laser pulse. (a) Sequence of PPM images of NSMT2. Presented are (left to right) -2.66 ps, -0.67 ps, 1.33 ps, 3.33 ps and 5.33 ps where time zero is the point at which the fs-e and laser pulses arrive at NSMT2 simultaneously. Horizontal and vertical lines indicate positions of (b) and (c) which are time-dependent slices through the dataset, and the black arrows indicate the evolution of charging of NSMT2. (b) (left) Part of the -2.67 ps spatial image below the white line, rotated clockwise by 90 degrees. (right) Sections through the time-dependent image dataset along the horizontal line in (a) as a function of laser-fs-e delay. Each slice is 9 pixels wide, corresponding to 50 nm. The flare indicated by the black arrows at positive delay is due to electron-electron scattering from apex of NSMT2 induced by strong-field electron ejection. (c) (left) As (b) but for the left-hand part of the -2.67 ps image, followed by (right) the temporal evolution of 9 pixel wide slice as function of laser-fs-e delay. In all images, unity on the colour scale corresponds to 10$^4$ electrons summed over 5 $\times$ 10$^5$ laser shots, and the scale bars in (b) and (c) are 1 micron.}
	\label{fig4}
\end{figure}

\section{Results - femtosecond electron microscopy of a nanoscale metal tip}
The separation between NSMT1 and NSMT2 was reduced to hundreds of microns. With no laser illumination on NSMT2, NSMT1 was translated from the calibrated magnification position producing a series of images as shown in figure \ref{fig1}(b). Typical magnifications were in excess of 10$^3$, which should elucidate the features on the apex of NSMT2, however are not sufficient to facilitate the observation of in-line holograms. 

With NSMT1 and NSMT2 laser illuminated, the delay between the arrival time of the fs-e pulse and the vertically polarized laser pulse at NSMT2 scanned. The NSMT1 to NSMT2 flight time prediction from the GPT model allowed constraints to be placed on the delay range. Furthermore two fast photodiodes (DET-10A, Thorlabs) recorded the laser pulses on the exit of the vacuum chamber giving an absolute measurement of the laser pulse arrival times, albeit at a resolution 5 orders of magnitude lower than the laser pulse durations. 

As shown in figure \ref{fig4}(a), as the delay between the horizontal and vertical polarized laser pulses is scanned over 8 picoseconds, the image of the apex of NSMT2 appears to expand, followed by the shank of NSMT2 blurring at longer delays. Laser illumination of NSMT2 causes low energy electron ejection from the apex just as with NSMT1. As this charge distribution is not strongly accelerated, it remains in the vicinity of NSMT2 and the trajectory of the passing fs-e pulse is distorted by electron-electron repulsion. The expansion and propagation of the dark region with time indicates regions of high charge density on the surface of the nanotip triggered by the laser pulse. These regions deflect passing electrons in the fs-e probe pulse expanding the shadow of the tip around the regions of highest charge density. 

Figure \ref{fig4}(a) reveals that the charge propagation appears to occur in stages. Between temporal overlap and approximately 1.3 ps there is a strong but localized increase in charge density concentrated at the very end of the tip, which for positions close to the apex has a rise time of approximately 1.5 ps. Between 1.3 and 2.3 ps this charge cloud begins to expand away from the apex, before finally, between 2.3 and 3.3 ps, the charge cloud begins to propagate down the taper of the tip. This occurs preferentially on the side of the tip from which the laser is incident. 

We elucidate the temporal behaviour by presenting horizontal (figure \ref{fig4}(b)) and vertical (figure \ref{fig4}(c)) sections through the time-varying image of NSMT2, at the positions indicated by the white lines in figure \ref{fig4}(a). These are shown in the right-hand panels of figure \ref{fig4}(b) and (c). In figure \ref{fig4}(b), the lower part of the left-hand most image in figure \ref{fig4}(a) is presented for spatial reference, and the left-hand panel in figure \ref{fig4}(b) is rotated clockwise 90 degrees. 

As the delay switches from fs-e then laser (negative delay) to laser then fs-e (positive delay), the time-dependent expansion of the apex of NSMT2 can be observed, the result of the onset of laser-induced charging, which then dissipates down the shank of NSMT2 over hundreds of picoseconds. The mechanics behind this process will be the subject of future work. Here, we only assume that the charging process can be as short in time as the laser pulse duration. 

It is interesting to note that the charge expansion observed at NSMT2 does not have reflection symmetry. In figures \ref{fig4}(a) and (c), the laser pulse arrives from the left, while in \ref{fig4}(b) it arrives from the top. It is suggested that the asymmetry of emission is then due to diffraction of the vertically polarized laser pulse by NSMT2, thus modifying the emission site shape.

\begin{figure}
	\includegraphics[width=0.5\linewidth]{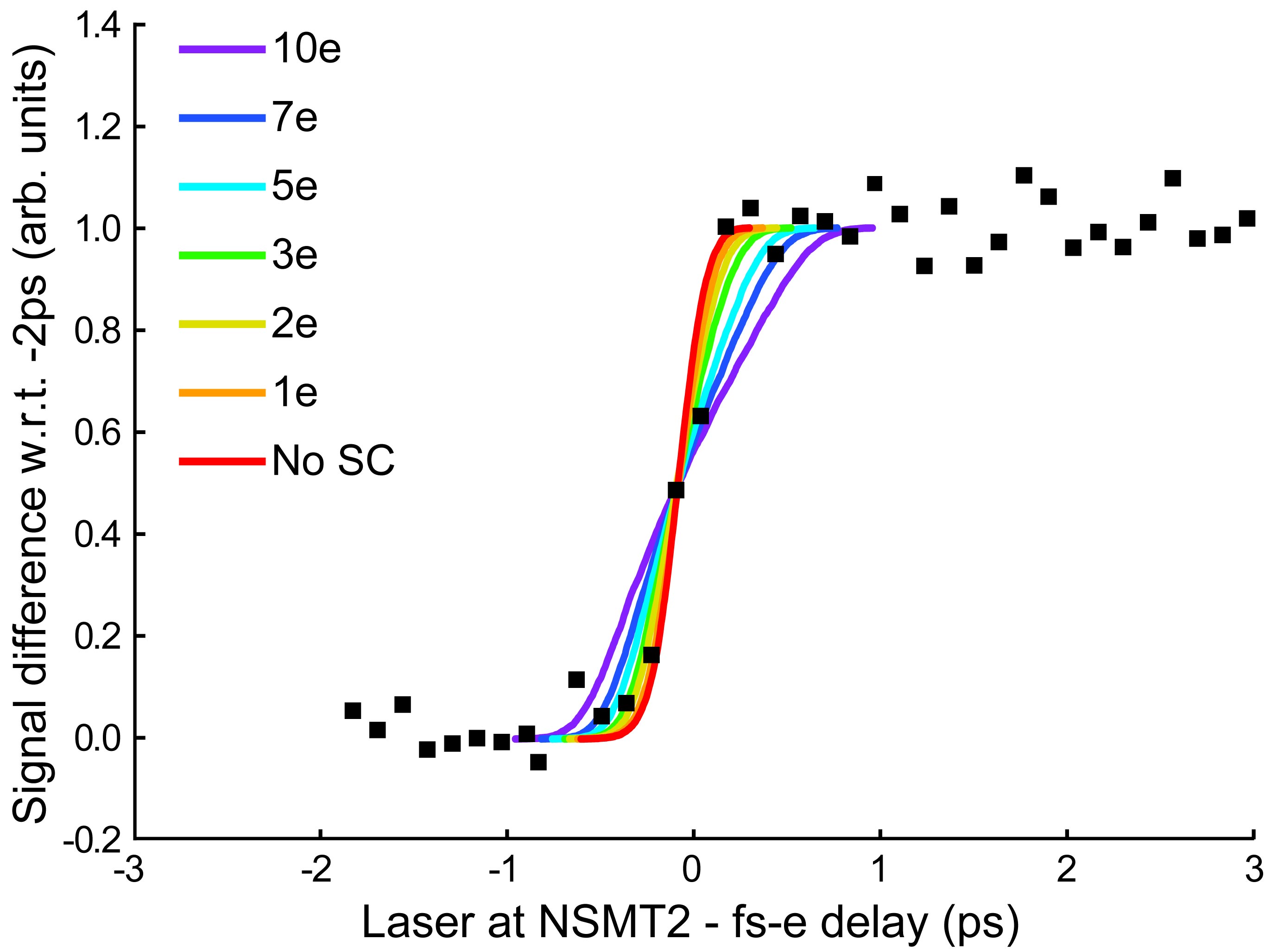}
	\caption{Comparison between the measured difference between large negative laser-fs-e pulse delay and subsequent fs-ePPM images (black squares), and the predicted cumulative electron flux at the plane of NSMT2 (coloured lines) as the average number of electrons per pulse is varied between zero electrons (No SC) and 10 electrons. The laser-induced charging of NSMT2 switches on rapidly but dissipates slowly, hence using the cumulative signal difference. The signal is found by taking a 9$\times$9 square of pixels on subsequent images and taking the difference between the comparable point on the -2 ps dataset. The comparison of these measurements to the GPT predictions allows the number of electrons per pulse to be inferred. }
	\label{fig5}
\end{figure}

\section{Results - number of electrons per fs-e pulse and temporal resolution}
The data in figure \ref{fig4} contains information about the laser-induced process around NSMT2, however we limit ourselves to the fastest behaviour at the apex. The rationale is that a process developing over a characteristic timescale shorter than the duration of the fs-e pulse gives metrological insight. Furthermore, other fs-e characterisations are not favourable in the current configuration. Looking for thermionic charging of the TEM grid as employed by Ropers and coworkers\cite{gulde} is limited to picoseconds. Ponderomotive electron scattering from counter-propagating laser pulses\cite{Hebe} requires pulse energies orders of magnitude higher than the NOPA is capable of generating. Electron streaking is too slow\cite{Kassier}. RF cavity deflection\cite{poormansxfel1} is not possible with the current configuration and is limited in terms of synchronization. The nonlinear response of NSMT2 to 30 fs laser illumination is therefore a good compromise: it is highly localized, of a comparable duration and is very well synchronized. 

The spatial region exhibiting the most rapid response is the apex of NSMT2 where the field enhancement is strongest therefore resulting in the highest instantaneous change in charge. In figure \ref{fig5} we show this temporal response, where the largest negative delay is taken as a zero point, and subsequent fs-e images of NSMT2 sequentially subtracted. We take a 9 $\times$ 9 pixel region around the point of fastest response, a compromise between statistical stability and suppressing the temporal response. In figure \ref{fig5} a rise-time below 1 ps is observed. Quantification of the electron flux using the method described above found between 0.8 and 12 electrons per pulse arrived at the detector within 2 SD of the mean. To refine this quantification, GPT was employed to predict the distribution of macro-particle arrival times at the plane of NSMT2. The time of release from the apex of NSMT1 Gaussian with a SD of 30 fs. As this surface is spherical, the flight time to the plane of NSMT2 is influenced by the trajectory, which along with the finite bandwidth of the fs-e pulse implies a duration greater than 30 fs. 

The laser-initiated enhancement at NSMT2 switches on in around ten fs, then takes picoseconds to redistribute and for the apex to neutralize. Direct comparison of the measured rise-time with the cumulative number of macro-particles arriving at a specific time represents how the fs-e pulse is distorted. The outcome of these calculations is shown in figure \ref{fig5} where the total charge of the fs-e pulse is varied between zero and ten electrons. With zero total charge, GPT still treats the pulse electrostatically but there is no space charge broadening. As the total charge in the pulse increases, a temporal broadening is predicted which is asymmetric around the temporal centre of the pulse. 

\begin{figure}
	\includegraphics[width=0.4\linewidth]{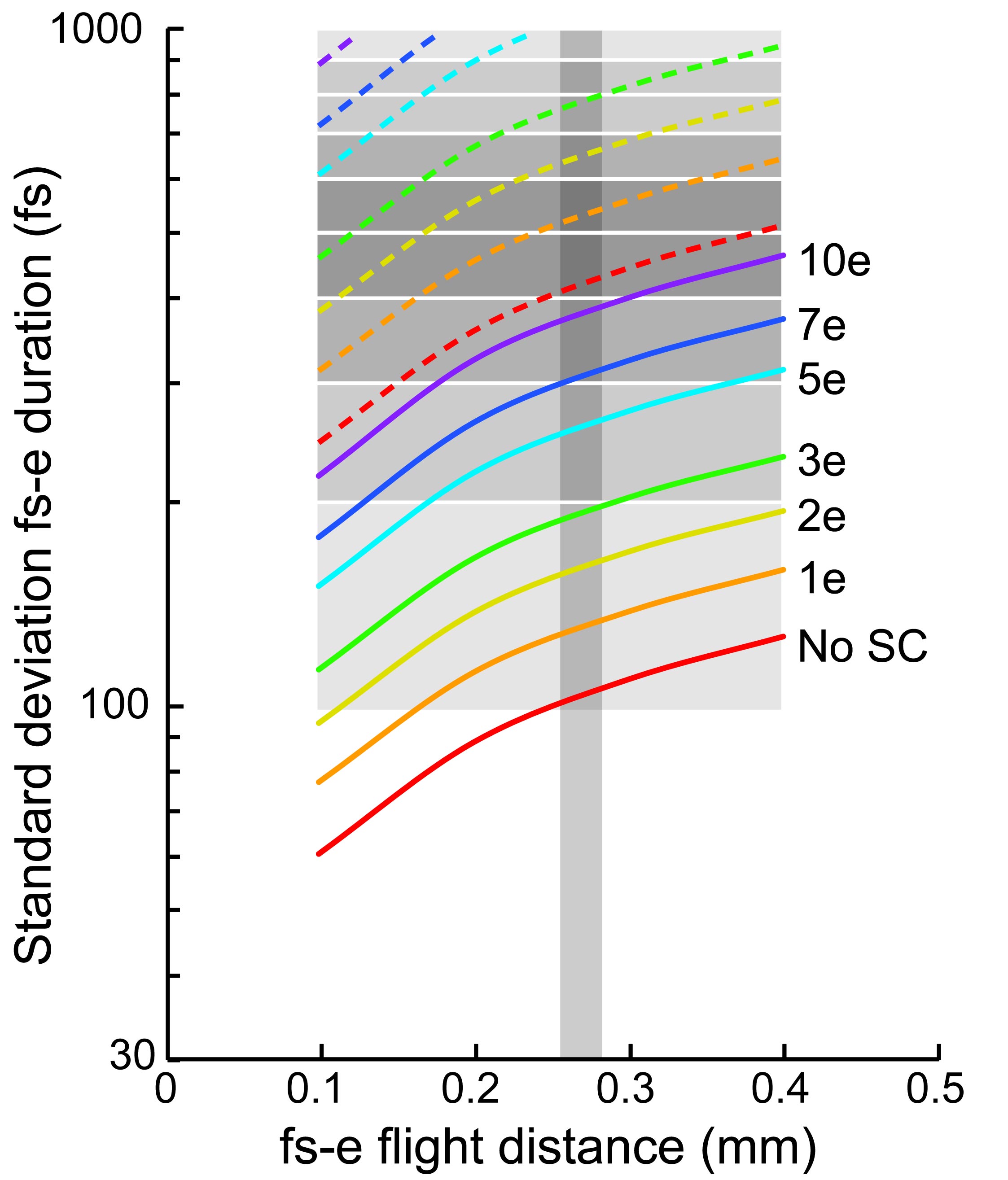}
	\caption{Femtosecond electron pulse stretching on propagation from NSMT1 including space charge, geometric effects and energy bandwidth. The solid lines are the predicted standard deviation of the fs-e pulse width as a function of time at different points along the flight path between NSMT1 and NSMT2. The SD duration of the space-charge free fs-e pulse increases along the flight path, a result of the geometric stretch and energy bandwidth. The contribution of space charge can also be observed as the average amount of total charge in the pulse is increased from no space charge (No SC) to 10 electrons per pulse. The dashed lines are the 2$\sigma$ width of the corresponding Gaussian distribution, i.e. 4$\times$SD. Experimental constraints are indicated by the grey regions. The vertical thin grey box indicates the position of NSMT2 with respect to NSMT1, and the horizontal grey boxes indicate an estimate of the rise time of the signal. The grey-level varies linearly with gradient of the signal, seen to be a maximum between 400 and 600 fs. The overlap between these regions and the dashed lines facilitates an estimate of the electron flux, which is most likely to be between less than one electron to two electrons per pulse. }   
	\label{fig6}
\end{figure}

\begin{figure}
	\includegraphics[width=0.6\linewidth]{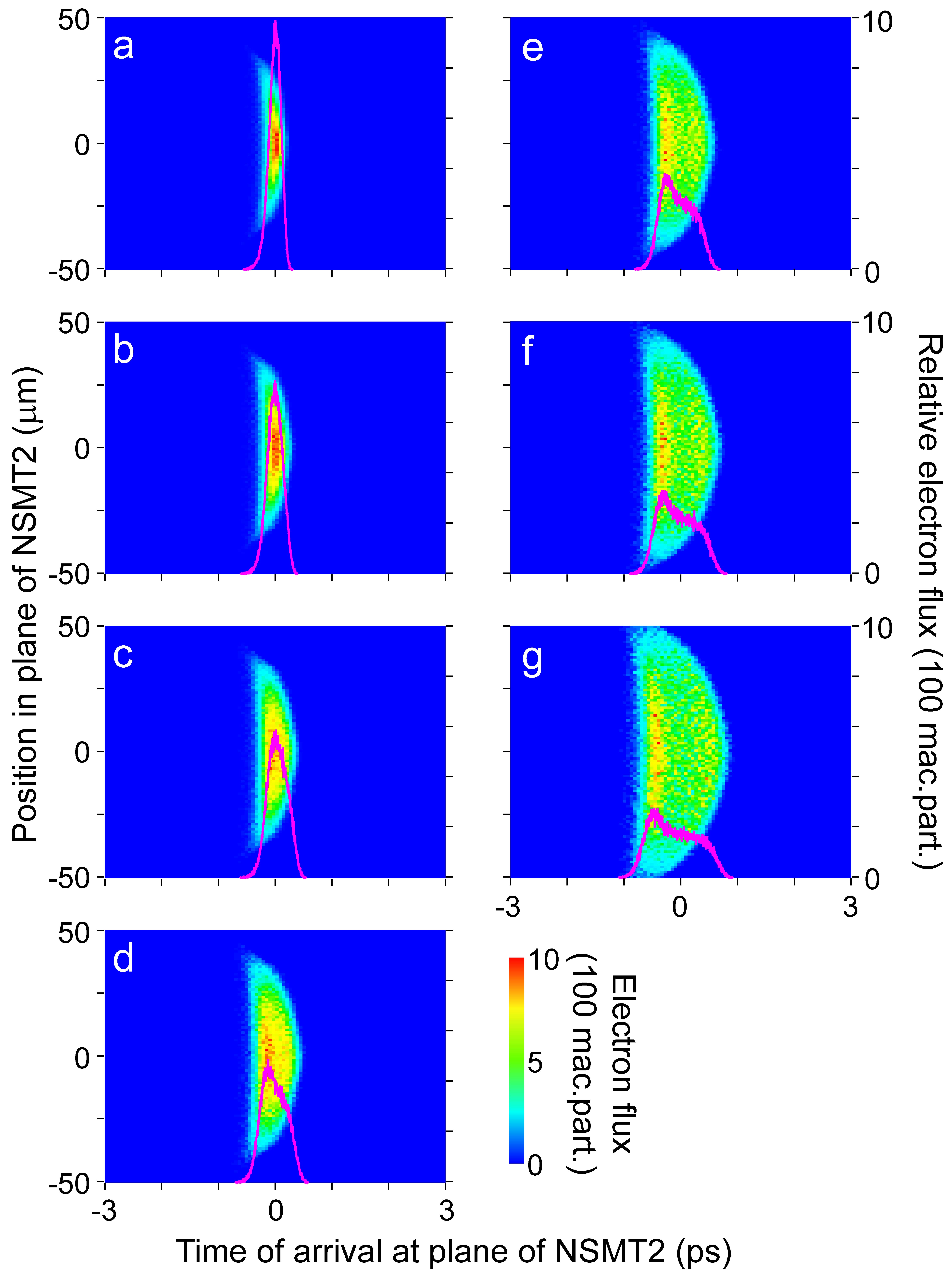}
	\caption{Electron pulse spatial distribution as a function of arrival time at the plane of NSMT2 (colour density plots) and corresponding histogram of the relative electron flux (magenta line). The total charge of the electron pulse is varied from (a) 0e (No SC), (b) 1e, (c) 2e, (d) 3e, (e) 5e, (f) 7e) and (g) 10e, with the total charge evenly distributed across the 50,000 GPT macro-particles used in the calculation. As the total charge of the pulse increases, a significant spatial and temporal spreading is found. The distribution best representing the results presented in figure $\ref{fig5}$ are (a), (b) and (c).}   
	\label{fig7}
\end{figure}

Comparing the modelled and measured rise-times in figure \ref{fig5}, it is apparent our fs-e pulse on average comprises fewer than ten electrons. Indeed, looking at the agreement between the data around the largest gradient as a function of time, we argue that an average charge per pulse between one and three electrons is the best fit to the experimental data. This is highlighted in figure \ref{fig6} where we show the SD fs-e duration as a function of propagation distance, indicating space charge broadening. Should it be possible to separate NSMT1 and NSMT2 by 100 $\mu$m or less, electron pulses well below 100 fs could be delivered to target on the condition of a few electrons per pulse. Two regions of constraint are indicated on figure \ref{fig6} connected to two experimental conditions. An estimate of the NSMT1 to NSMT2 distance is included (vertical grey bar), as is the rise-time of the observations (graded horizontal bars). The overlap of these regions then indicates the most likely number of electrons per pulse after scaling the predicted SD by a factor of 4. This corresponds to the 95$\%$ confidence width of the distribution and better represents the maximal resolvable influence of the passing fs-e pulse. 

Comparing our gauge calibration with that of figures \ref{fig5} and \ref{fig6} allows deduction of an upper bound of 7 electrons per laser shot, which although somewhat rough, provides an order of magnitude agreement with a single electron pulse. Looking at figure \ref{fig6}, it is apparent that the femtosecond electron pulses employed in the present work are best described by an average total charge of approximately one electron per pulse. Such pulses are expected to be nearly free from space charge effects, and when coupled to DC or RF acceleration methods along with well-synchronozed compression methods, will allow few-femtosecond or even attosecond duration pulses of electrons to be produced.

In figure \ref{fig7} we show the predicted temporal and spatial distribution of the electron pulse in the plane of NSMT2. Increasing the total charge disperses the pulse in space and time while keeping the centre of charge at the same time. Figure \ref{fig7}(a-c) represent our best estimate of the distribution of our fs-e pulse at NSMT2. The influence of the geometric stretch causes the wavefront curvature, hence manipulation of this wavefront could shorten the pulse, beneficial to higher temporal resolution studies. 

The simulations show that the effect of space-charge is drastic, with a 3 electron bunch displaying a FWHM approximately twice that of a single electron bunch. The simulations predict that bunches of the order of 1-10 electron population remains relatively short, with a 10 electron bunch approximately 1 picosecond in duration at the sample for the geometry considered.

\begin{figure}
	\includegraphics[width=0.4\linewidth]{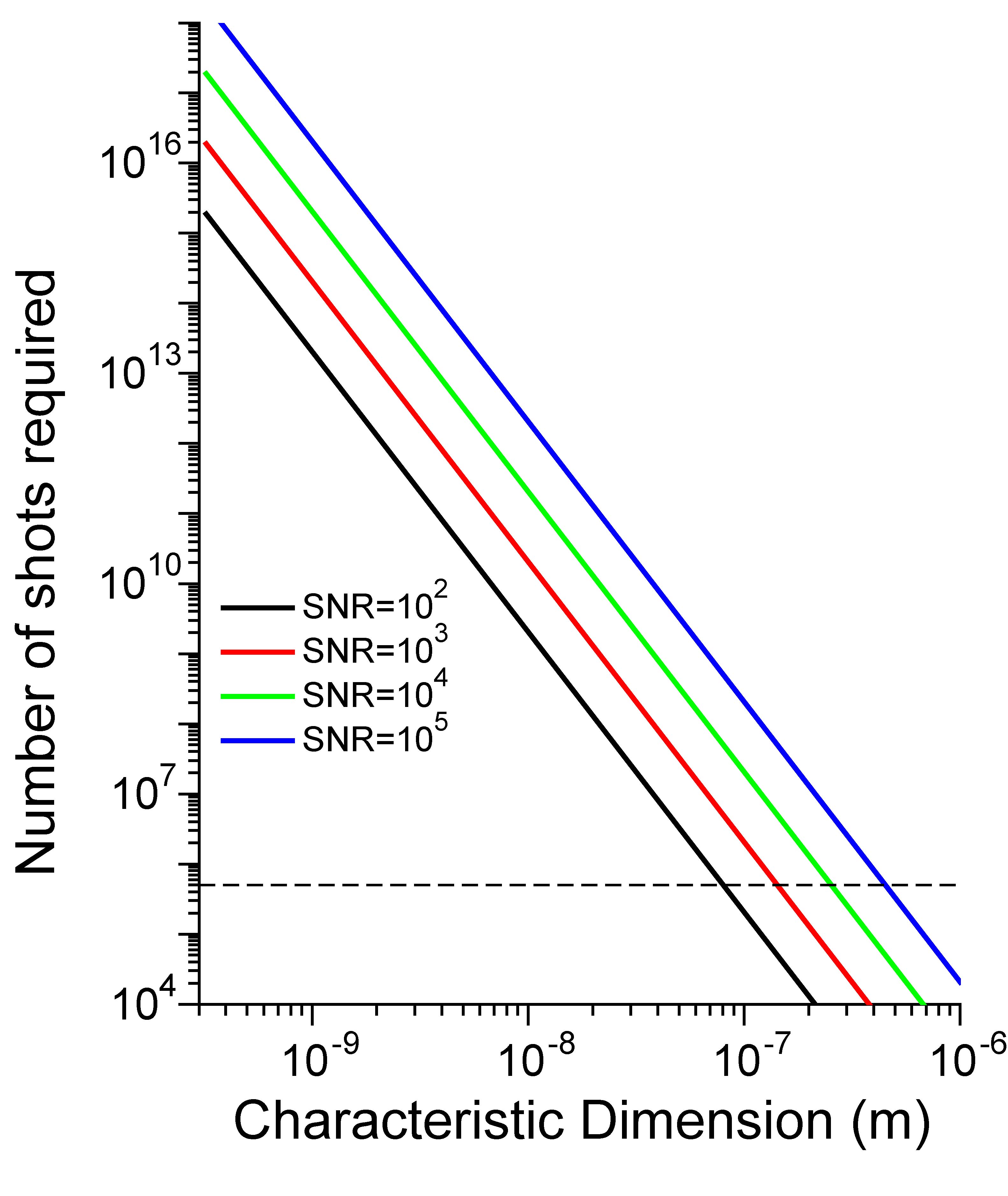}
	\caption{Quantification of the number of laser shots required for real-space imaging with 800 nm, 30 fs laser illumination at 10 GW/cm$^2$ and 150 eV electron pulses at one electron per pulse for a range of signal-to-noise ratios. The experimental results in figure 5 correspond to an SNR of approximately 100 and were collected with 5 $\times$ 10$^{5}$ shots, indicated by the horizontal dashed line. The dimension of the unit cell of tungsten is 3.24 \AA.} 
	\label{fig8}
\end{figure}

\section{Discussion of spatial resolution}
In the experimental results presented in figure 4, we estimate the spatial resolution to be 100 nm, which is the minimum observable change in feature position within the signal to noise ratio (SNR) of the data. By considering the mechanics of photon absorption and electron scattering we quantify the number of real-space imaging events required to produce an image with a specific SNR. This is intended as an estimate of magnification limitations of the current experimental configuration and as a cross-check on claims of resolution.

The absorption coefficient for tungsten at 800 nm\cite{Rakic, Weaver} is 4.4338 $\times$ 10$^{-5}$ cm$^{-1}$, corresponding to an absorption cross section of 6.86 $\times$ 10$^{-18}$ cm$^2$.  The peak laser intensity at the apex of NSMT2 is assumed to be of the order 10 GW/cm$^2$, above which we would expect significant degradation due to plasma formation. Given the density of 19.3 g/cm$^3$, we find the number of photons arriving per unit area (A), hence via the cross section find the probability of excitation via two photon absorption, which is assumed to be resonant. 

For an electron of 150 eV, the inelastic mean free path is 4.5 \AA, so we consider electron scattering from the area A, and only consider one atomic layer, with thickness 3.24 \AA. The total cross section for elastic scattering of 150 eV electrons\cite{Salvat} is 8.88 $\times$ 10$^{-16}$ cm$^2$, and at the plane of NSMT2, we assume our electron flux of one electron per shot passes through a disc of radius 30 $\mu$m, which is reasonable considering our earlier estimations of source size. This flux is then converted into the probability that an electron in area A is scattered.

Combining the photon absorption and electron scattering probabilities gives the probability of a real-space event that will contribute to the time dependent signal observed. Then, for a given SNR, the total number of events required for area A to produce an observable time-dependent signal can be found, as shown in figure 8 for a range of characteristic dimensions. The data in figure 5 at a delay greater than zero has a SNR of approximately 100, corresponding to the black line in figure 8, collected with 5 $\times$ 10$^5$ laser shots, indicated by the horizontal dashed line. This is in favourable agreement with the resolution estimate from figure 4. 

It should be noted that this calculation is performed for a potential of 150 V applied over 260 $\mu$m. The resulting low electric field gradient was used so as to completely suppress DC field emission. It is anticipated that through experimental advances such as the use of grating-coupled NSMTs and micro-machined electrodes, the source-sample distance will be further decreased while increasing the field gradient. This will reduce the fs-e pulse duration while increasing the electron flux per unit area such that unit cell resolution will be approached.

To consider thermal effects, we assume all photons in a cone with base radius 50 nm, and height 1 $\mu$m receives 10 GW/cm$^2$, which is clearly a worse case scenario. At a laser repetition rate of 50 kHz we would expect to see an increase in temperature of approximately 1 mK, therefore significantly higher repetition rates could be tolerated before thermionic emission or other thermal effects played any part in the dynamics at the apex of NSMT2.

\section{Conclusions}
Point projection electron microscopy of a tungsten nanotip has been performed with sub-micron scale spatial resolution using femtosecond laser emitted photoelectrons. By tuning the emission such that an average of a single electron is released per laser pulse, space-charge broadening is overcome allowing images with a temporal rise-time of below 500 fs associated with a standard deviation in time of 120 fs to be obtained. This has been verified by tracking the propagation of charge along a tungsten nanotip induced by a femtosecond laser pulse, and is to the author's knowledge the fastest combined temporal and spatial response recorded, especially with single electron pulses used for real-space imaging. For single electron ultrafast diffraction measurements, Bragg spots have been tracked with characteristic time-scale of around 700 fs, reported recently\cite{baum1} albeit with uncompressed electron pulses. Our measurement highlights the potential of femtosecond electron point-projection microscopy for high resolution measurements of charge dynamics in both space and time. The temporal resolution demonstrated here expands the range of ultrafast processes that can be investigated with pulsed electron sources, particularly those susceptible to damage by transmission of keV to MeV pulses. 

The correlation between the timescales of observed features and the best temporal resolution predicted by the simulations indicates that the single electron per laser shot emission regime is being exploited and that sub-100 fs temporal resolution is possible for probing electrons with sub-keV energies. We are investigating the ideal laser parameters for a range of fs-ePPM energies and anticipate a further increase in temporal resolution will be possible. An analysis of photon absorption followed by electron scattering from an area of the target gives a good agreement with our observed spatial resolution, which while low in the current configuration, has significant promise for imaging localised charging and dynamic electric and magnetic fields, with implications for future ICT devices.

Installing a NSMT in a femtosecond electron microscopy or diffraction apparatus is straightforward, therefore the presented method could be adapted as a general fs-e pulse characterisation technique. This technique is best suited for low total charge electron pulses (less than hundreds of electrons) at repetition rates of tens of kilohertz and higher, rather than low (few kHz and lower) high charge (10$^5$ electrons) systems. 

\begin{acknowledgments}
These experiments were carried out with UFL1 from the Engineering and Physical Sciences Research Council (EPSRC, UK) Laser Loan Pool. We are very grateful to Dr. Ian Clark, Central Laser Facility, STFC Rutherford Appleton Laboratory (UK) for his support of this project during his management of the Laser Loan Pool. ARB acknowledges Postgraduate Studentship support from EPSRC, UK, and CWBM from a College of Science, Swansea University Postgraduate Research Studentship. 
\end{acknowledgments}

\bibliography{bibliography}{}

\end{document}